\documentclass{pasj00}
\draft

\begin{document}
\SetRunningHead{Ojha et al.}{V1647 Orionis : A New Look at McNeil's Nebula}
\Received {yyyy/mm/dd}
\Accepted {yyyy/mm/dd}

\title{{\bf V1647 Orionis} (IRAS 05436-0007) : A New Look at McNeil's Nebula}

\author{D.K. \textsc{Ojha}}
\affil{Tata Institute of Fundamental Research, Mumbai (Bombay) - 400 005, India}
\email{ojha@tifr.res.in}
\and
\author{N. \textsc{Kusakabe}, M. \textsc{Tamura}, Y. \textsc{Nakajima}, 
and M. \textsc{Fukagawa}}
\affil{National Astronomical Observatory of Japan, Mitaka, Tokyo 181-8588, Japan}
\and
\author{K. \textsc{Sugitani}}
\affil{Institute of Natural Sciences, Nagoya City University, Mizuho-ku,
Nagoya 467-8501, Japan}
\and
\author{D. \textsc{Baba}, D. \textsc{Kato}, M. \textsc{Kurita}, 
C. \textsc{Nagashima}, T. \textsc{Nagayama}, T. \textsc{Nagata}, 
and S. \textsc{Sato}}
\affil{Department of Astrophysics, Faculty of Sciences,
Nagoya University, Chikusa, Nagoya 464-8602, Japan}


\maketitle

\begin{abstract}
We present a study of the newly discovered McNeil's nebula in Orion using the 
JHK$_s$-band simultaneous observations with the near-infrared (NIR) camera 
SIRIUS on the IRSF 1.4m telescope. The cometary infrared nebula is clearly 
seen extending toward north and south from the NIR source 
{\bf (V1647 Orionis)} that 
illuminates McNeil's nebula. The compact nebula has an apparent diameter of 
about 70\arcsec. The nebula is blue (bright in J) and has a cavity structure 
with two rims extending toward north-east and north-west. The north-east rim 
is brighter and sharp, while the north-west rim is diffuse. The north-east rim 
can be traced out to $\sim$ 40\arcsec~from the location of the NIR source. In 
contrast, no cavity structure is seen toward the south, although diffuse nebula
is extended out to $\sim$ 20\arcsec. New NIR photometric data show a 
significant variation in the magnitudes ($>$ 0.15 mag) of the source of 
McNeil's nebula within a period of one week, that is possibly under the phase 
of {\bf eruptive} variables like FUors or EXors.
\end{abstract}

\section{Introduction}

On the night of 2004 January 23, Jay McNeil discovered a new reflection nebula 
in the L1630 cloud in Orion (McNeil 2004). His amazing discovery is now 
recognized as a newly visible reflection nebula surrounding a 
{\bf young star} -- McNeil's nebula. \mbox{McNeil's} new object 
{\bf (V1647 Ori)} 
seems to be a faint optical counterpart to the infrared source catalogued as 
\mbox{IRAS 05436-0007} that has gone into outburst, producing a large 
reflection nebulosity (Reipurth \& Aspin 2004; Abrah\'am et al. 2004;
Brice\~no et al. 2004; Vacca, Cushing \& Simon 2004; Andrews, 
Rothberg \& Simon 2004; {\bf Walter et al. 2004}). Little is known about the 
source. The discovery was announced by B. Reipurth in the Star Formation 
Newsletter 
No. 136.\footnote{http://www.ifa.hawaii.edu/$\sim$reipurth/newsletter.htm}
Reipurth \& Aspin (2004) reported that NIR images taken with the Gemini 8m 
telescope show that the object has brightened by about 3 magnitudes in JHK, 
relative to the 1998 2MASS measurements. Gemini spectra reveal strong features 
of CO and Br$\gamma$ in emission in the IR; in the optical, H$\alpha$ shows 
emission with a P Cygni profile (Reipurth \& Aspin 2004).   

The McNeil's source at the apex of the nebula was also observed to brighten 
dramatically in X-ray and optical wavelengths. The X-ray data from 
{\it Chandra} gives strong evidence that the probable cause of the outburst is 
the sudden infall of gas onto the surface of the star from an orbiting disk of 
gas (Kastner et al. 2004). 

In this letter we present the infrared morphology of McNeil's nebula
as well as infrared monitoring of its central source, especially,
in a short (about one week) timescale. Both of these pieces of information
will be useful to understand the nature of this enigmatic source.
We carried out simultaneous JHK$_s$-band photometry of the McNeil's nebula 
during five nights from 
2004 Feb 17 to March 14 and report a significant variation in the source
brightness of McNeil's nebula within a period of one week. 
These observations also show a new look at McNeil's nebula in NIR 
wavelengths. In \S 2 we present the details of observations and data
reduction procedures. \S 3 deals with the results and discussion on
short time variability of the star that {\bf illuminates} McNeil's nebula and
describe the details of infrared nebula. We then summarize our
conclusions in \S 4.    

\section{Observations and data reduction}

The imaging observations of the McNeil's nebula region in the NIR wavelengths 
J ($\lambda$ = 1.25 $\mu$m), H ($\lambda$ = 1.65 $\mu$m), and 
K$_s$ ($\lambda$ = 2.15 $\mu$m) were obtained on 2004 February 17 (22:00 UT),
23 (18:36 UT), 25 (17:55 UT), 29 (17:48 UT), and 2004 March 14 (17:51 UT) with 
the Infrared Survey Facility (IRSF) 1.4m telescope situated at the South 
African Astronomical Observatory, Sutherland and SIRIUS
(Simultaneous three-color InfraRed Imager for Unbiased Surveys),
equipped with three 1024$\times$1024 HgCdTe arrays. The field of view in each
band is $\sim$ 7\arcmin.7 $\times$ 7\arcmin.7, with a pixel scale of
0\arcsec.45.
Further details of the instrument are given in Nagashima et al. (1999) and
Nagayama et al. (2003).

{\bf For the photometry of the bright NIR source, 
%counterpart of the
%nebula that is positionally coincident with the IRAS source (IRAS 05436-0007), 
we obtained 10 dithered exposures} of the target centered at
($\alpha$, $\delta$)$_{2000}$ = ($05^h46^m14^s.0$, 
\mbox{-00$^{\circ}05^{\arcmin}40^{\arcsec}.1$)}, each 1s long to avoid the
saturation, simultaneously for each band during all four nights.
Total on-target
integration times in each of the bands were 10s for the photometry of bright 
NIR counterpart on all four nights.  

We also obtained 10 dithered exposures 
on 2004 Feb 17 each 30s long, 15 dithered exposures on 2004 Feb 23 and 25 
each 10s long, and 45 dithered exposures each 20s long
on 2004 Feb 29 and March 14 of the target to study the morphological 
details of {\bf the reflection nebula}. Total on-target
integration times in each of the bands were 
%10s for the photometry of bright NIR counterpart on all four nights, and 
300s, 150s, 150s, 900s, and {\bf 900s 
%for the morphological study of McNeil's nebula 
on 2004 Feb 17, 23, 25, 29, and March 14, respectively}. 

The observations
on 2004 Feb 23, 25, and 29 were done under good photometric sky conditions.
The NIR images were obtained at high airmass ($\sim$ 1.9) and in poor
seeing condition (FWHM $>$ 2\arcsec)
%in J, H and K$_s$-bands) 
on 2004 Feb 17 and the sky condition was not photometric during 2004 March 14.
Therefore we use these observations for the study of the morphological details 
only. The average seeing sizes (FWHMs) in all the bands were 2.2\arcsec, 
1.6\arcsec, 0.9\arcsec, 1.4\arcsec, and 0.9\arcsec on 2004 Feb 17, 23, 25, 
29, and March 14, respectively
during the observations. The observations were made at airmasses between
1.2 and 1.4 during 2004 Feb 23 - 29. Dark frames and twilight flats were 
obtained at the beginning
and end of the observations. The photometric calibration was obtained by
observing the standard star 9116 in the faint NIR standard star catalogue of
Persson et al. (1998) at air masses closest to the target observations.    

Data reduction was done using the pipeline software based on NOAO's 
IRAF\footnote{IRAF is distributed by the
National Optical Astronomy Observatories, which are operated by the
Association of Universities for Research in Astronomy, Inc., under contract
to the National Science Foundation.} package tasks. Twilight flat-fielding
and sky subtraction with a median sky frame were applied. Identification
and photometry of point sources were performed by using the DAOFIND and
DAOPHOT packages in IRAF, respectively. We used an aperture radius of 
16 pixels ($\sim$ 7\arcsec) for the photometry of the McNeil's source. 
The local sky was evaluated in an annulus with an inner radius of 64 pixels 
and a width of 12 pixels. 

\section{Results and Discussion}

\subsection{IR morphology {\bf of McNeil's} nebula}

The composite color images are constructed from the SIRIUS J, H, and K$_s$-band
images (J represented in blue, H in green, and K$_s$ in red) obtained on
different seeing (FWHMs) conditions and are shown
in Fig. 1. The lower right panel (Fig. 1d) shows the RHK$_s$ (R: blue,
H:green, K$_s$: red) composite color image {\bf of McNeil's} nebula after
adjusting the spatial resolutions. R-band image was acquired using 
GMOS-N camera on the Gemini north telescope (Reipurth \& Aspin 2004).
The cometary infrared nebula is clearly seen extending toward north from the 
NIR source (see Figs. 1a, 1b and 1c). {\bf It has a strong morphological 
resemblance to a ring-shaped nebula, RNO 54, which is associated with an
H$\alpha$-emission star (Goodrich 1987)}.  
{\bf The McNeil's nebula} is blue (bright in J) and has 
a cavity structure with two rims extending toward north-east and north-west. 
The north-east rim is brighter and sharp, while the north-west rim is diffuse. 
The north-east rim can
be traced out to $\sim$ 40\arcsec~or $\sim$ 16000 AU at a surface
brightness of K$_s$ $\sim$ 18.2 mag arcsec$^{-2}$ (where we assume
the distance to L1630 is 400 pc; Anthony-Twarog 1982). 
%Counter cavity structure is not seen toward the south in our image.
This kind of morphology can be probably seen 
when the eruption has cleared out some of the dust and gas
in the envelope 
surrounding the young star, allowing light to escape and illuminate a 
cone-shaped cavity carved out by previous eruptions 
%into the gas 
(Reipurth \& Aspin 2004).
Counter cavity structure is not seen toward the south in our image.

In Fig. 2 we show the contour diagram of the region around {\bf V1647 Ori} as 
seen in our deep K$_s$-band image obtained on 2004 Feb 29.
%and March 14, 2004.
The star is surrounded by a compact reflection nebulosity that has a 
{\bf curved tail (north-east rim) characteristic} 
(see also Reipurth \& Aspin 2004). There is a southward 
continuation of the extended nebula that is shown in the PSF subtracted image 
of Fig. 3. 
This is confirmed by comparing the PSF of the NIR central source which is the
counterpart of the IRAS source with the PSF 
of the star located about 30\arcsec~south-west of the IRAS source (Fig. 3).  
{\bf We thus find that the compact nebula that encloses all the nebular 
structure including the two rims has an apparent diameter of about 
70\arcsec.} 

The IR/optical comparison of the region surrounding McNeil's nebula 
(see Figs. 1d and 4) shows that the position of the bright infrared source is 
coincident with that of an optical source at the apex of the nebula that is 
illuminating the fan-shaped cloud of gas, or nebula (Reipurth \& Aspin 2004).
It is interesting to compare the morphology of the optical nebula 
(see Figs. 1d, 4 and Fig. 1 of Reipurth \& Aspin 2004 ) and our IR nebula 
(Figs. 1 - 4). The optical nebula is more widely and
predominantly extended to the north, while the IR nebula is
relatively confined but definitely extended to the south, too.
In fact, the IR nebula well matches the reddest parts of
the optical nebula. 
%and the 
The bluest portion of the optical nebula
is not seen in IR including the blue knot HH 22 to the north-east
and the blue patch to the north. There is a clear color gradient 
from north to south, as well as from inside to outside the cavity.
The gradient is very steep near the central star and the 
southern extension of the nebula is seen only at IR. 
This large color gradient
and the sudden absence of optical nebula to the south is
suggestive of a large scale disk-like structure (or envelope) 
surrounding the central source
which hides the southern nebula. This is often the case for
the YSOs associated with a cometary nebula (e.g., Tamura et al. 1991).

\subsection{Short timescale variability of the central IR source}

NIR photometry {\bf of McNeil's} source obtained on three nights is listed in 
Table 1. A comparison of the JHK$_s$ magnitudes of {\bf V1647 Ori} 
%NIR counterpart of IRAS 05436-0007 
shows that the source has varied over the time of observation between 2MASS
(Oct 1998) and IRSF (Feb 2004). Reipurth \& Aspin (2004) have first shown 
that the object has become brighter by $\sim$ 3 mag in 
J, H, and K$_s$ bands (Table 1). Our photometry is more accurate 
($\Delta$m $\sim$ 0.03) than their measurements, although the values
are consistent within 0.1 mag error of the Gemini data. 
We also see a significant variation in 
the source brightness ($\Delta$J = 0.16 mag, $\Delta$H = 0.18 mag, and 
$\Delta$K$_s$ = 0.22 mag) during a period of one 
week (2004 Feb 23 - 29; Table 1). 
Since the difference in magnitudes is much 
larger than the photometric errors, we believe that this is real. To examine 
this, we also did the photometry of three bright reference stars 
in our IR FOV (see Table 2).
% $\sim$ 80\arcsec~north-east to IRAS 05436-0007 (Table 2). 
As seen in Table 2, we do not see a large variation in the 
magnitudes of the reference stars. This further proves that
the brightness variation in {\bf V1647 Ori} during a period of one week is 
real. It also shows that the
variability is larger at longer wavelengths than shorter ones
and the brightness decreased by about 0.2 mag at J, H and K$_s$-bands 
during 2004 Feb 23 - 29 (Table 1).
%We detect no significant color changes (J-H \& H-K$_s$) over a period of 
%one week (Feb 23 - 29, 2004; Table 1).
Walter et al. (2004) have followed the source in optical and NIR wavelengths
over the 25 nights (2004 Feb 11 - May 7) and also found 
a general decline in brightness of about 0.3-0.4 magnitudes during these
87 days based on the differential light curves. They presented the 
differential photometry {\bf of McNeil's} source in NIR wavelengths with 
respect 
to the only available comparison star (2MASS 05461162-0006279) in their
smaller IR FOV. 
%however nothing else is known about the nature of 
%the comparison star (variable or steady).
We also examined this 2MASS source and found that the brightness of the 
2MASS source is fairly constant during a period of one week (see Table 2). 
Therefore, our observations
give absolute value to the McNeil's source variability and are more
reliable.     

Fig. 5 shows a J-H/H-K color-color (CC) diagram. For the purpose of 
plotting the SIRIUS data in CC diagram, we have converted them into
the California Institute of Technology (CIT) system using the color
transformations between the SIRIUS and CIT systems 
(Nakajima et al., in prep)
\footnote{available at
http://www.z.phys.nagoya-u.ac.jp/$\sim$sirius/about/color$_{-}$e.html},
that have been obtained by observing several of the red standard stars of
Persson et al. (1998).   
We notice the changes of the
infrared colors from before (1998) to after the eruption (2004).  
The star has moved precisely along a reddening vector due to the intrinsic
brightening of the star between six years as well as between one week 
%(Reipurth \& Aspin 2004) 
and is moving toward the tip of the locus
of classical T Tauri stars (Meyer et al. 1997). This suggests that
the central source has an NIR color similar to a dereddened T Tauri star. 
It is to be noted that the seeing is almost the same for the data points
1 and 3 but different for point 2 (Fig. 5 and Table 1). Therefore, the
comparison between 1 and 3 is more reliable. 
The J-H and H-K values
indicate that the circumstellar matter of A$_V$ $\sim$ 8 mag was cleared
probably in the eruption, 
%definitely 
in less than six years since the 2MASS observations. There is still   
a significant amount (A$_V$ $\sim$ 6 mag) of 
circumstellar matter if its original colors are assumed to be those
of T Tauri stars.

Variability in low-mass
YSOs is known (e.g., Hartmann 1998). The total luminosity depends on the mass
accretion rate and low mass YSOs like FUor and EXor types of objects 
have shown variability of several magnitudes in the optical and NIR
wavelengths ($\Delta$m $\sim$ 3-4 mag) 
%and NIR wavelengths 
(Bell et al. 1995, and references therein). It is possible that we 
have witnessed an FU Orionis kind of behavior as the source has about the
right amplitude (Reipurth \& Aspin 2004). On the other hand EXor eruptions,
which may occur repeatedly in a given star, have amplitudes that can be 
comparable to those of FUors but with much shorter durations 
(Reipurth \& Aspin 2004). Our data show a significant variation in
the source brightness within a period of one week (Table 1). 
So far, there have been no infrared data showing such significant 
variation within a period of one week or so.
Further 
photometric monitoring of the object is required to settle the class of eruptive 
variables to which source of McNeil's nebula belongs.      
 
\section{Conclusions}

In this letter we have presented the short time variability and the new 
look of the McNeil's nebula in NIR wavelengths. We   
carried out deep simultaneous NIR observations of the McNeil's nebula 
during five nights. From the analysis of NIR images we derive the 
following conclusions : 

1) The McNeil's nebula has a cavity structure with two rims as seen in the
NIR images. This might have been produced by the eruption that has cleared out
a part of the envelope surrounding the young star.
%some of the dust and gas surrounding the young star. 

2) The IR/optical comparison of the McNeil's nebula
shows that the optical nebula is more widely and
predominantly extended to the north, while the IR nebula is
relatively confined but definitely extended to the south, too.
This color gradient of the nebula is most likely due to
the presence of a large scale disk around the young star.

3) The NIR photometric data show a significant 
variation in the source brightness \mbox{($>$ 0.15 magnitudes)} 
within a period of one week. 
%We find that the brightness of the source has decreased by about 
%0.2 mag at K over a period of one week.    

4) Based on the NIR photometry, a post outburst extinction (A$_V$) to
McNeil's source appears to be about 6 mag, if the intrinsic JHK$_s$ 
colors of the central star are the same as that of a T Tauri star. 

%5) Although it is not clear whether McNeil's source falls in FUor 
%or in an EXor category yet these 
%NIR observations of the present outburst will provide additional knowledge on
%the source. 

%\Acknowledgments

We thank the referee, Prof. George Herbig, for a most thorough 
reading of this paper and several useful comments and
suggestions, which greatly improved the scientific content of the paper.
We thank the staff of the South African Astronomical Observatory for their
kind support during the observations. 
DKO was supported by the Japan Society for the Promotion of Science
(JSPS) through a fellowship during which most of this work was done. 
MT acknowledges support by Grant-in-Aid 12309010 from the Ministry of
Education, Culture, Sports, Science, and Technology.
We thank Colin Aspin and Bo Reipurth for providing us with the FITS images of
their Gemini optical bands {\bf of McNeil's} nebula.

\clearpage

\begin{table}
\caption{JHK photometry {\bf of McNeil's source (V1647 Ori)}}
%\footnotesize
\begin{tabular}{ccccccccc}
\\
\hline
\hline
Date (UT)       & JD &    J         &   H           &   K$_s$   & K$^\prime$  & K  & FWHM$^a$ & Telescope \\
\hline
1998 Oct 7      & -- & 14.74$\pm$.03 & 12.16$\pm$.03 & 10.27$\pm$.02 & -- & -- & & 2MASS \\
2004 Feb 3      & -- & 11.1$\pm$.1 & 9.0$\pm$.1 & -- & 7.4$\pm$.1 & -- & 0.5\arcsec & Gemini$^b$\\
2004 Feb 11     & -- & 10.79$\pm$.01 & 8.83$\pm$.01 & -- & -- & 7.72$\pm$.01 & -- & USNO$^d$\\  
2004 Feb 23$^c$ & 2453058.784 & 10.74$\pm$.05 & 8.93$\pm$.01 & 7.40$\pm$.02 & -- & -- & 1.6\arcsec & IRSF\\
2004 Feb 25$^c$ & 2453060.756 & 10.79$\pm$.05 & 8.95$\pm$.01 & 7.47$\pm$.01 & -- & -- & 0.9\arcsec & IRSF\\
2004 Feb 29$^c$ & 2453064.782 & 10.90$\pm$.06 & 9.11$\pm$.01 & 7.62$\pm$.02 & -- & -- & 1.4\arcsec & IRSF\\
2004 Apr 12     & -- & 10.94$\pm$.02 & 9.06$\pm$.02 & -- & -- & 7.59$\pm$.02 & -- & USNO$^d$ \\
%Mar 14, 2004$^a$ & 10.71$\pm$.06 & 9.65$\pm$.02 & 7.55$\pm$.04 & IRSF\\
\hline
\end{tabular}
\\
$^a$ Measured FWHM of a standard star. This is a measure of the seeing.

$^b$ Reipurth \& Aspin (2004): photometry is obtained in an aperture with radius 0.9\arcsec

$^c$ IRSF photometry is obtained in an aperture with radius 7.2\arcsec.

$^d$ McGehee et al. (2004)
\end{table}

\begin{table}
\caption{JHK photometry of the reference stars at IRSF}
%\footnotesize
\begin{tabular}{ccccc}
\\
\hline
\hline
Reference star        &       J        &       H      &     K$_s$     &  Date (UT)\\
\hline
2MASS05461946-0005199 & 10.81$\pm$.07 &  9.73$\pm$.02 &  8.96$\pm$.07 & 2004 Feb 23\\
                      & 10.77$\pm$.05 &  9.70$\pm$.02 &  8.97$\pm$.04 & 2004 Feb 29\\
\\
2MASS05460575-0002396 & 12.34$\pm$.02 & 11.27$\pm$.01 & 10.82$\pm$.02 & 2004 Feb 23\\
                      & 12.35$\pm$.02 & 11.29$\pm$.01 & 10.87$\pm$.01 & 2004 Feb 29\\
\\
2MASS05461162-0006279 & 13.81$\pm$.07 & 12.26$\pm$.04 & 11.32$\pm$.05 & 2004 Feb 23\\
                      & 13.80$\pm$.07 & 12.27$\pm$.02 & 11.33$\pm$.02 & 2004 Feb 29\\
\hline
\end{tabular}
\end{table}

\clearpage

\begin{figure}
\begin{center}
%\FigureFile(175mm,175mm){f1.eps}
\end{center}
\caption{JHK$_s$ three-color composite images {\bf of McNeil's nebula} (J: blue, 
H: green, K$_s$: red) obtained with SIRIUS mounted on the IRSF 1.4m telescope 
on 2004 Feb 23 [a], 
%(seeing $\sim$ 1.6\arcsec), 
Feb 25 [b] 
%(seeing $\sim$ 0.9\arcsec) 
and Feb 29 [c], 
%(seeing $\sim$ 1.4\arcsec), 
respectively. 
%The total integration times are 150s, 150s and 900s on Feb 23, 25 and 29, 
%respectively. 
d) RHK$_s$ three-color optical/IR composite image (R: blue, H: green, 
K$_s$: red) after adjusting the spatial resolutions. R-band image was 
acquired with Gemini on 2004 Feb 14 (Reipurth \& Aspin 2004) and H and 
K$_s$-band images were obtained with SIRIUS on 2004 Feb 25. North is up and 
east is to the left.
\label{fig1}}
\end{figure}

\clearpage

\begin{figure}
\begin{center}
%\FigureFile(175mm,175mm){f2.eps}
\end{center}
\caption{{\bf A compact nebula (the nebular structure including the two rims)} 
is seen around the illuminating star of McNeil's
nebula in \mbox{K$_s$-band} image obtained on 2004 Feb 29.
% and March 14, 2004, respectively.
%The total on source integration time is 900 s.
The lowest contour and the contour interval are 18.2 mag arcsec$^{-2}$ and
0.3 mag arcsec$^{-2}$, respectively.
%for both images. 
The abscissa and the ordinate are in J2000 epoch.
\label{fig2}}
\end{figure}

\clearpage

\begin{figure}
\begin{center}
%\FigureFile(175mm,175mm){f3.eps}
\end{center}
\caption{The PSF subtracted image of McNeil's nebula region in H-band 
obtained on 2004 Feb 23 displayed in a logarithmic intensity scale.
North is up and east is to the left. The locations of the NIR central source 
{\bf (V1647 Ori)} which is the counterpart of the IRAS source 
and the star $\sim$ 30\arcsec~south-west of {\bf V1647 Ori} 
%the IRAS source 
are marked 
by filled triangles in the H-band image (see the text). 
\label{fig3}}
\end{figure}

\clearpage

\begin{figure}
\begin{center}
%\FigureFile(175mm,175mm){f4.eps}
\end{center}
\caption{IR and optical images of the region surrounding McNeil's nebula.
SIRIUS K$_s$-band image (grey color) obtained on 2004 Feb 29 is overlaid with 
contours from the optical R-band image obtained with Gemini 
(Reipurth \& Aspin 2004). The abscissa and the ordinate are in J2000 epoch.
\label{fig4}}
\end{figure}

\clearpage

\begin{figure}
\begin{center}
%\FigureFile(175mm,175mm){f5.eps}
\end{center}
\caption{J-H/H-K color-color diagram showing the location of {\bf V1647 Ori}
%counterpart of IRAS 05436-0007 
as observed with 2MASS (filled circle) on
1998 Oct 7; with IRSF (filled squares) on 2004 Feb 23 [1], Feb 25 [2] and 
Feb 29 [3]; with Gemini (filled triangle) on 2004 Feb 3; and with USNO 
(asterisks) on 2004 Feb 11 and Apr 12.
The sequences of field dwarfs (solid curve) and giants (thick dashed curve)
are from Bessell \& Brett (1988). The dotted line represents the locus of 
T-Tauri stars (Meyer et al. 1997). Dashed straight lines represent the
reddening vectors (Rieke \& Lebofsky 1985). The crosses on the dashed lines
are separated by A$_V$ = 5 mag. 
\label{fig5}}  
\end{figure}

\end{document}